\documentclass[10pt,a4paper]{article}
\usepackage[dvips]{graphicx,color}
\usepackage{amsmath,amssymb}
\usepackage{latexsym}

\usepackage{ulem}

\begin{document}
\def\dfrac#1#2{{\displaystyle\frac{#1}{#2}}}
\def\cfrac#1#2{\dfrac{\mathstrut #1}{#2}}
\newcommand{\Slash}[1]{\ooalign{\hfil/\hfil\crcr$#1$}}
\title{Constraints on 
a vacuum energy
from both SNIa and CMB temperature observations }
\author{Riou  Nakamura$^{\dagger}$, E. P. Berni Ann Thushari$^{\dagger}$,  Mikio Ikeda$^{\dagger}$,  and
 Masa-aki Hashimoto$^{\dagger}$ \\
${}^{\dagger}$ {Department of Physics, Kyushu University}, \\
{6-10-1, Hakozaki, Higashi-ku, Fukuoka-city, 812-8581, Japan}
}

\maketitle
\begin{abstract}
We investigate the cosmic thermal evolution with a
  vacuum energy which decays into photon at the low-redshift.  
We assume that the vacuum energy is  a function of the scale factor that increases 
toward the early universe.
We put on the constraints  using  recent observations of both type Ia supernovae (SNIa)  by Union-2
 compilation and  the cosmic microwave background (CMB) temperature  
at the range of the redshift $0.01 < z <  3$. 
From SNIa,  we find that the effects of a decaying
 vacuum energy on the  cosmic expansion rate should be  very small but could be
 possible for $z < 1.5$.
On the other hand, we obtain the severe constraints for parameters from the
CMB temperature observations.  Although the temperature can be
still lower than the case of the standard cosmological model, 
it should only affect the thermal evolution at the early epoch. 
\end{abstract}

\section{Introduction}

One of the biggest cosmological mysteries is the accelerating cosmic expansion which
was discovered by the observations of distant type Ia supernovae (SNIa) starting from more than 10 years ago.
For the origin of the accelerating expansion, the following possibilities have been proposed:
the modified gravity such as $f(R)$ gravity\cite{fr-grav},  brane-world cosmology~\cite{Umezu2006},
inhomogeneous cosmology~\cite{Inhomogeneous}, and existence of unknown  energy called as dark energy
which is equivalent to a cosmic fluid with a negative pressure.

It is strongly suggested that the dark energy amounts to 70\% of the total energy
density of the universe from the astronomical observations such as 
 SNIa \cite{rf:Perlmutter,Union-2}, the anisotropy of cosmic microwave background (CMB) ~\cite{QUaD,WMAP7},
 and the baryon acoustic oscillation~\cite{BAO}. 
Although there are various theoretical models of dark energy  (details are shown in a review \cite{DEreview}), 
its physical nature is still unknown.
{
Considering  above-referenced observations, we can not exclude the
constant $\Lambda$ term} which is  the most simple model of dark energy.

The interacting dark energy with CMB photon
has been discussed, where
a decaying vacuum into photon is related to a primordial
light-element abundances~\cite{Freese1986}  and the
CMB intensity~\cite{Overduin1993}.
From the point of the thermodynamical evolution in the universe, a
decaying vacuum energy modifies the
temperature-redshift relation~\cite{Lima1995,Jetzer2011}.
This modification affects the cosmic thermal evolution after a hydrogen
recombination~\cite{rf:Kimura}, such as formation of  molecules~\cite{rf:Hashimoto,rf:Puy} and the 
first star~\cite{rf:Hashimoto}.
In addition to this model, the angular power spectrum of CMB could be
also modified~\cite{Nakamura2008}.
In the previous analysis,  
thermal history at the higher redshift has been studied. However, as seen in \cite{rf:Kimura,rf:Hashimoto},
the time variation of temperature at low redshift differs from the model without a decaying-vacuum

In the present study, we update the observational consistency of a
vacuum energy (hereafter we denote as $\Lambda$) coupled with CMB photon.
 Hereafter we call this model D$\Lambda$CDM (model).
To examine the consistency with observations at low redshift,  we focus
on the cosmic evolution comparing both the type Ia
supernovae and the redshift dependence of CMB temperatures.
This is because the time dependence of the CMB temperature for a D$\Lambda$CDM differs from CDM model
with a constant $\Lambda$ term of which hereafter we call S$\Lambda$CDM (model).
In \S 2 formulation of  D$\Lambda$CDM is reviewed. 
In \S 3 the $m-z$ relation is investigated for D$\Lambda$CDM  in a flat
universe. 
Parameters inherent in this model are
constrained from the CMB temperature observations in \S 4. 
 Concluding remarks are given in
\S 5.

\section{Dynamics of the decaying $\Lambda$ model}
The Einstein's field equation is  written as follows~(e.g. \cite{Weinberg2008}): 
\begin{equation}
R_{\mu\nu}-\frac{1}{2}g_{\mu\nu}R  = {8\pi G}{}T^{}_{\mu\nu},
\label {p13_ee}
\end{equation}
where $G$ is the gravitational constant and $T^{}_{\mu\nu}$ is the
energy momentum tensor. 
If we assume the perfect fluid, $T^{}_{\mu\nu}$ is written as follows:
\[
 T^{}_{\mu\nu} = {\rm diag}{\left( -\rho, p, p, p\right)}.
\]
Here $\rho$ and $p$  are the energy density  and the pressure, respectively.
Note that we choose the unit of $c=1$.

  The equation of motion is obtained with use of the
  Friedmann-Robertson-Walker metric of the homogeneous and isotropic principle:
\begin{equation}
ds^{2}=-dt^{2}+a\left(t\right)^{2}\left[{\frac{{d}r^{2}}{1-kr^{2}}+r^{2}d\theta^{2}+r^{2}\sin^{2}{\theta}d\phi^{2}}\right], 
\label{aaa}
\end{equation}
where $a(t)$ is the scale factor and $k$ is the specific curvature constant. This leads to the Friedmann equations: 
\begin{equation}
H^{2}\equiv  \left(\frac{\dot{a}}{a}\right)^{2}=\frac{8\pi
 G}{3}\rho-\frac{k}{a^{2}}. 
\label{p14_d}
   \end{equation}

From the conservation's law of energy, 
we   can obtain the equation of the energy density:
\begin{equation}
 \dot{\rho} = -3\left( 1 + w \right)\frac{\dot{a}}{a}\rho,
\label{eq:rhodot}
\end{equation}
where $w$ are the coefficients of the equation of state, $p=w\rho$, and defined by 
\begin{equation}
 w \equiv \begin{cases}
\frac{1}{3} ~~~\text{for photon and  neutrino}, \\
0 ~~~\text{for baryon and cold dark matter},\\
-1 ~~~\text{for vacuum energy}.
\end{cases}
\notag 
\end{equation}


We take  the component of the energy density as 
\begin{equation}
 \rho = \rho^{}_{\gamma} + \rho^{}_{\nu} + \rho^{}_{m} +
  \rho^{}_{\Lambda},
\notag 
\end{equation}
 where $\gamma, \nu, m$, and $\Lambda$  indicate photon, neutrino,
 non-relativistic matter
 (baryon plus cold dark matter), and the vacuum,  respectively.
Here we neglect the energy contribution of $\Lambda$ on other components
except for photon: the $\rho^{}_{\nu}$ and $\rho^{}_{m}$ evolves as $\rho^{}_{\nu}\propto a^{-4}$
and $\rho^{}_{m}\propto a^{-3}$, respectively. From
 eq.~\eqref{eq:rhodot},  the equation of the
 photon coupled with $\Lambda$ is written as 
\begin{equation}
 \dot{\rho}^{}_{\gamma} + 4H\rho^{}_{\gamma} = -\dot{\rho}^{}_{\Lambda}.
\label{eq:rhodot_gamma}
\end{equation}
Let us define the parameter of the energy density as follows:
\begin{equation}
 \Omega^{}_{i} \equiv \frac{\rho^{}_{i}}{\rho^{}_{cr}} = \frac{8\pi G}{3H^{2}_0}\rho^{}_{i},
\label{eq:energy_parameter}
\end{equation}
 where $\rho^{}_{cr}$ is the critical density defined by the present Hubble
constant $H^{}_0$.
We can rewrite eq.\eqref{eq:rhodot_gamma}  
as follows~(details also see \cite{rf:Puy,Nakamura2008}):
\label{p14_dn} \\
\begin{equation}
\frac{d \Omega_{\gamma}}{d a}+4\frac{
 \Omega_{\gamma}}{a}=-\frac{d \Omega_{\Lambda}}{da}.
\label{p14_g} 
\end{equation}

 

   Models with time dependent $\Lambda$-term have
   been studied as summarized in \cite{rf:Overduin, Sahni1999}. 
We adopt the energy of the vacuum  which varies
   with the scale factor \cite{rf:Kimura,rf:Hashimoto,rf:Puy,Nakamura2008}:
   \begin{equation}
\Omega^{}_{\Lambda}(a)=\Omega^{}_{\Lambda1}+\Omega^{}_{\Lambda2}a^{-m},
\label{p14_wb}
 \end{equation}
where $\Omega_{\Lambda1}, \Omega_{\Lambda2}$, and $m$ are constants. 
Note that the present value of $\Omega^{}_{\Lambda}$ is expressed by 
$\Omega^{}_{\Lambda0} =\Omega^{}_{\Lambda1}+\Omega^{}_{\Lambda2}$.

Integrating eq.~\eqref{p14_g} with  eq.~\eqref{p14_wb}, we obtain the
photon energy density as a function of $a$ ~\cite{rf:Puy}:
\begin{equation}
\Omega^{}_{\gamma} =
 \begin{cases} 
  {\left[ \Omega^{}_{\gamma0}+\alpha\left( a^{4-m}_{}-1 \right)
  \right]}a^{-4}_{}  & \left( m\ne 4 \right), \\ 
  \left( \Omega^{}_{\gamma0}+4{\Omega^{}_{\Lambda 2}}
 \ln{a}\right)a^{-4}_{} & \left( m=4 \right),
\end{cases}
\label{eq:rhogamma}
\end{equation}
where $\Omega^{}_{\gamma 0}=2.471\times 10^{-5}_{}h^{-2}_{}( T^{}_{\gamma 0}/2.725 {\rm ~K})^4$
is the present photon energy density, 
$h$ is the normalized Hubble constant ($H^{}_{0} = 100 \, h$ km/sec/Mpc)
, $T^{}_{\gamma0}$ is the CMB temperature at the present epoch.
We define $\alpha$ as $\alpha=m\Omega^{}_{\Lambda2}/(4-m)$.


Since we assume the flat geometry $(k=0)$ in this work, we can write the condition of $\Omega$s as follows:
\[
 \Omega^{}_{m0} + \Omega^{}_{\Lambda1} + \Omega^{}_{\Lambda2} = 1.
\]
 From now on, we adopt the present cosmological parameters: 
Hubble constant~
$h=0.738$~\cite{Riess2011},
 and 
the present density parameter of matter $\Omega_{m}= 0.2735$~ \cite{Jarosik2011}.
The present temperature is 
$T^{}_{\gamma0} = 2.725$~K observed by COBE~\cite{COBE1999}. 
In our study, we search  the parameter regions in the range of $\Omega^{}_{\Lambda2}$ and $m$:
$10^{-7}_{}<\Omega^{}_{\Lambda2} < 10_{}^{-2}$ and $0<m<4$.
The range of $m$ has already been obtained from the previous analysis~\cite{Nakamura2008}.

{
Here we note the range of $\Omega^{}_{\Lambda2}$ in the present analysis.
Kimura et al.~\cite{rf:Kimura} found that the radiation temperature (i.e., the energy density of 
radiation) in D$\Lambda$CDM model becomes numerically negative at a
point of $a<1$ when $m$ and/or 
$\Omega_{\Lambda2}$ is too large.
To avoid this unreasonable situation, i.e. $T^{}_{\gamma}<0$, Nakamura et al.~\cite{Nakamura2008} 
imposed 
the limit of $\Omega_{\Lambda2}$ and $m$ as follows:
\begin{align}
\alpha &< \Omega_{\gamma0}~~~(m<4), \notag \\
\Omega^{}_{\Lambda2} &< \Omega^{}_{\gamma0}/92~~~(m=4). \label{eq:condition_oml2}
\end{align}
From the condition \eqref{eq:condition_oml2}, we obtain 
the limit of $\Omega^{}_{\Lambda2}$ as $4.9\times10^{-7}$ for
$m=4$. 
We can consider that the effect of $\Omega^{}_{\Lambda2}$ is negligible
when $\Omega^{}_{\Lambda2}$ is smaller than $10^{-7}$.
}


\section{SNIa constraints}

The cosmic distance measures depend sensitively on the spatial curvature
and expansion dynamics of the models. Therefore the magnitude-redshift
relation for distant standard candles are proposed  
to constrain the cosmological parameters.

To calculate the effects of the comic expansion at low-$z$, 
we calculate the theoretical distance modules,
\begin{equation}
  \mu_{th}=m-M=5\log{d_L}+25,
\label{mu}
 \end{equation} 
where $m, M$, and $d_L$ are the apparent and absolute
magnitudes, and  the luminosity distance, respectively.
Here $d_L$ is related to the radial distance $r$ in the metric as follows~\cite{Weinberg2008}:
\begin{equation}
d_{L}=\left(1+z \right) r.
\label{magni}
\end{equation}
We can obtain $r$ as follows:
\[
 r = \int{\frac{dt}{a(t)}} = \frac{1}{H^{}_0}\int{\frac{dz}{E(z)}}, 
\]
where $z$ is the redshift defined by $a=1/(1+z)$.
Now $E(z)$ is defined as
\begin{equation}
E\left( z \right)^{2} =\left( H\left( z \right)/H_{0}\right)^{2}\\
= \Omega_{m0}\left(1+z \right)^3
+\Omega_{\Lambda1} + \Omega^{}_{\Lambda2}\left( 1+z\right)^m_{}  .
\label{mult}
\end{equation}
Then, Eq.~\eqref{magni} is rewritten as follows,
\begin{equation}
 d_L = \frac{\left( 1 + z \right)}{H_0} \int{\frac{dz}{E(z)}}.
\label{eq:dL-2}
\end{equation}
As a consequence, the theoretical distance modules are calculated by Eqs.~\eqref{mu}
and \eqref{eq:dL-2}.

\begin{figure}
   \centering
\includegraphics[width=.80\linewidth,keepaspectratio]{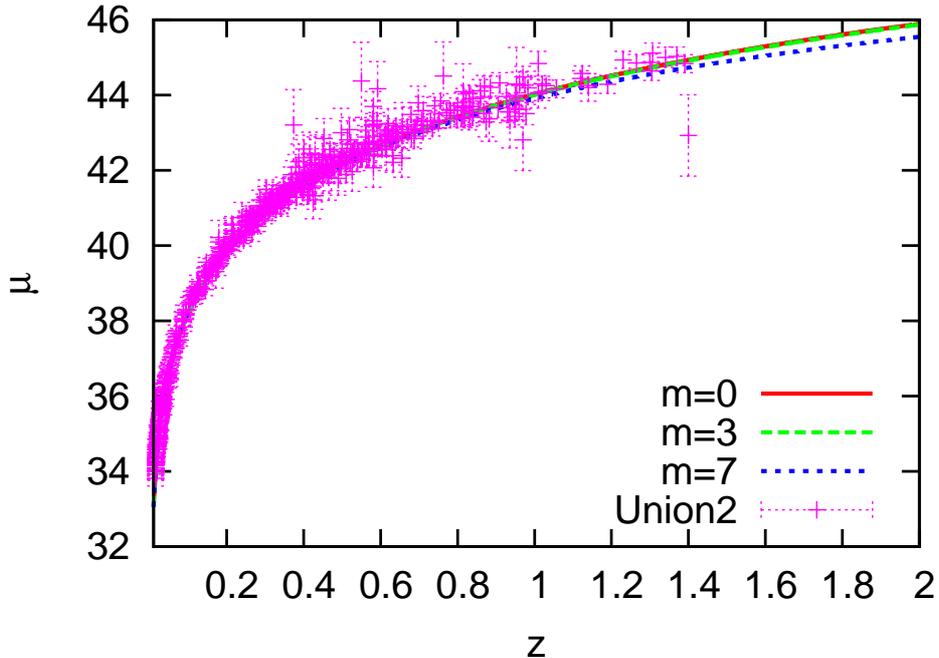}
     \caption{Illustration of the magnitude-redshift relation in
 S$\Lambda$CDM model ($m=0$) and  D$\Lambda$CDM with
 $\Omega_{\Lambda2}=10^{-2}$ compared with SNIa observations (cross-shaped ones with error-bars).
Note that the theoretical curve for $m=3$  has no difference with that
of $m=0$.}
        \label{m-zm}
   \end{figure}

\begin{figure}[tb]
\begin{center}
 \includegraphics[width=.70\linewidth,keepaspectratio]{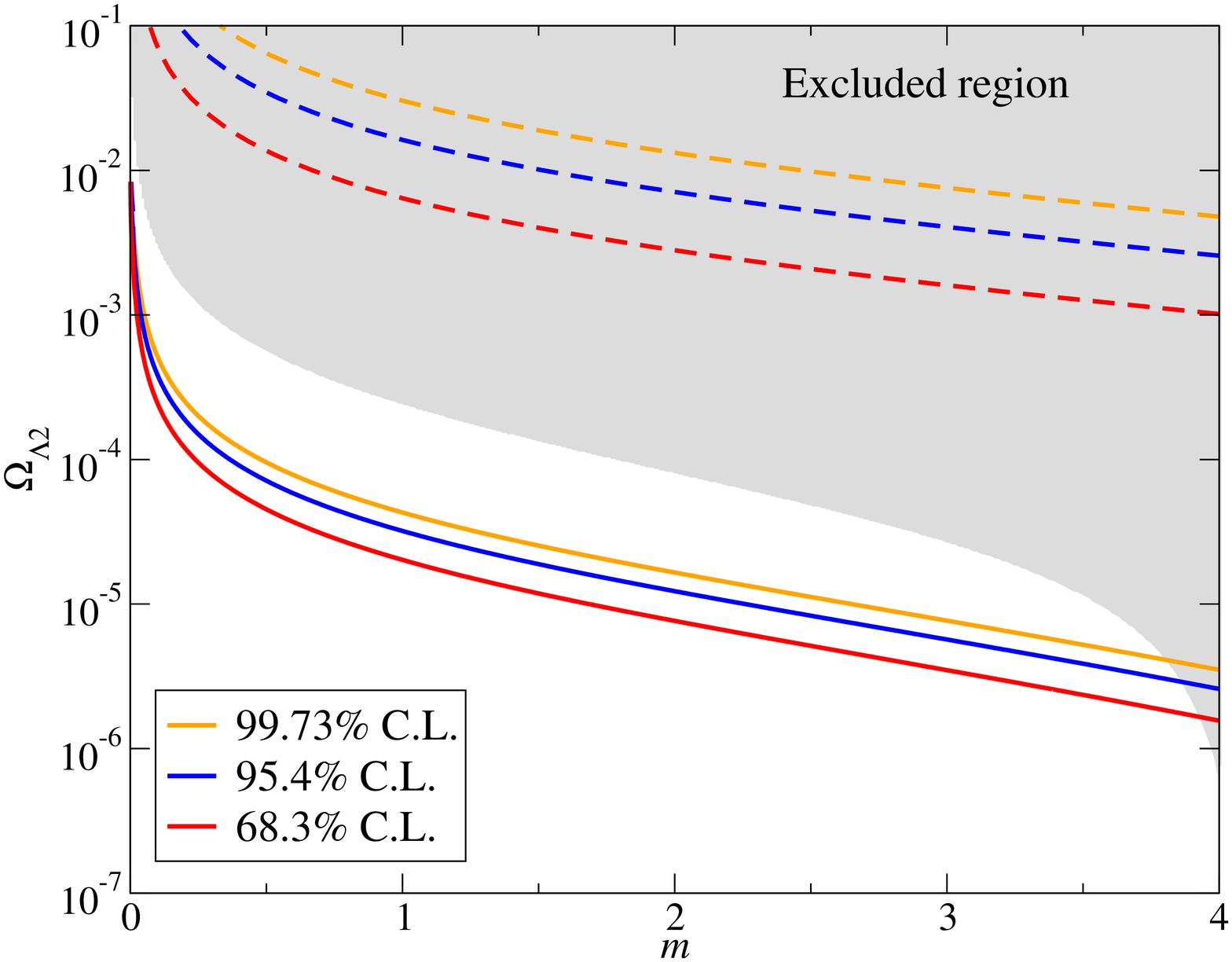}
\caption{Constraints on the $m-\Omega_{\Lambda2}$ plane from the
recent observations of CMB.  The lines indicate upper bounds of $1, 2$, and
 $3\sigma$ confidence level. The dashed- and solid-lines are the limits deduced from
 SNIa and CMB temperature observations. The shaded region shows the excluded
 region obtained from Ref.~\cite{Nakamura2008}.}
\label{fig:snIa_cntr}
\end{center}
\end{figure}

Recently, the Supernovae Cosmology Project (SCP) collaboration released
their Union2 sample of 557 SNIa data \cite{Union-2}. The
Union2 compilation is the largest published and spectroscopically  confirmed
SNIa sample. The observations are in the redshift range of $0.01<z<2$.
For the SNIa data set, we evaluate $\chi^{2}$ value of the distance modules, 
\begin{equation}
\chi^{2}=\sum_{i=1}^{N}\frac{\left(\mu_{th,i}-\mu_{obs,i}\right)^{2}}{\sigma^{2}_{\mu_{obs,i}}+\sigma^{2}_{v,i}},
 \label{ggg}
\end{equation}
where $\mu_{obs,i}$ and $\mu_{th,i}$ are the observed and the
theoretical values of the distance modules. $\sigma_{v}$ is the dispersion
in the redshift due to the peculiar velocity $v$ is written as,
\begin{equation}\label{new}
\sigma_{v}=\left( \frac{v}{c} \frac{d\mu_{th}}{dz}\right).
\end{equation}
We adopt $v = 300~\rm{km~sec^{-1}}$~\cite{rf:Kessler}.
The total number of the united sample $N$ is $557$ for the present analysis. 

 We confirm that  $\Omega_{\Lambda {2}}$ has done some contribution to change the
 $m-z$ relation. When $\Omega_{\Lambda2}$ increases,
 the expansion rate in the universe decreases, which is similar to the
 behavior of the matter dominant universe in the Friedmann model. 

  Figure \ref{m-zm} indicates the relation between the distance modules and
 the redshift  against the
 SNIa observations in terms of S$\Lambda$CDM and D$\Lambda$CDM with
 several values of $m$ for a fixed value of
 $\Omega_{\Lambda_2}=10^{-2}$.  
As the value of $m$-value increases, $\mu$ tends to decrease because of the
 increasing cosmic expansion rate.
We recognize that $m$ is not seriously
 effective to change the $\mu-z$ relation. 
Even if we chose the larger value of $m>4$,  the  effect is still small.

Figure \ref{fig:snIa_cntr} shows the allowed parameter region due to the
$\chi^2$ fitting of Eq.~\eqref{ggg} in D$\Lambda$CDM  from SNIa constraints. 
We obtain the minimum value of $\chi^{2}=677$ (the reduced $\chi^{2}_r$
is defined to be
 $\chi^{2}_{r}=\chi^2/N\simeq 1.215$ where $N$ is the degree of freedom).
The best fit parameter 
region of $\Omega_{\Lambda2}$ for $m<4$ is  investigated as
$\Omega_{\Lambda2}<4.5\times 10^{-3}$ at $1\sigma$ confidence level~(C.L) . 
Since the parameter region is very loose compared with the previous
analysis~\cite{rf:Puy,Nakamura2008}, 
this result imply that the
decaying-$\Lambda$ has minor effects on the expansion rate  at $z<2$.

\section{Temperature constraints}

\begin{table}
\caption{Observational temperatures obtained from molecular excitation levels and S-Z effects.}
\label{tab:temp_obs}
\begin{tabular}[t]{ccc}
\hline\hline
   $z_{obs}$  &  $T$ [K]  &   Reference \\
\hline\hline
$1.776$  &  $7.4\pm 0.8$ & CI~\cite{Songaila1994} \\
\hline
$1.9731$ &  $7.9\pm 1.0$  & CI~\cite{Ge1997}  \\
\hline
$2.3371$ &  $6.0 < T$ [K] $< 14.0$ & CI~\cite{Srianand2000} \\
\hline
$3.025$  &  $12.1^{+1.7}_{-3.2}$ & C+~\cite{Molaro2002} \\
\hline
$1.77654$ &  $7.2\pm 0.8$ &  CI~\cite{Cui2005} \\
\hline
 $2.4184$ & $9.15\pm0.72$ &   CO~\cite{Srianand2008} \\
\hline
 $2.6896$ & $10.5^{+0.8}_{-0.6}$ & CO~\cite{Noterdaeme2010} \\
\hline
 $1.7293$ & $7.5^{+1.2}_{-1.6}$ &  CO~\cite{Noterdaeme2011} \\
 $1.7738$ & $7.8^{+0.6}_{-0.7}$ & \\
 $2.0377$ & $8.6^{+1.0}_{-1.1}$ & \\
\hline
$0.203$  &  $3.377^{+0.101}_{-0.102}$ &  S-Z~\cite{Battistelli2003} \\
$0.0231$ &  $2.789^{+0.080}_{-0.065}$ &  \\
\hline
 $0.023$	&    $2.72\pm 0.10$ &  S-Z~\cite{Luzzi2009} \\
 $0.152$	&   $2.90\pm 0.17$  &  \\
 $0.183$  &   $2.95\pm0.27$   &  \\
 $0.20$    &   $2.74\pm 0.28$  & \\	
 $0.202$	&   $3.36\pm 0.2$   & \\
 $0.216$	&   $3.85\pm 0.64$  & \\	
 $0.232$  &   $3.51\pm 0.25$  & \\
 $0.252$	&   $3.39\pm 0.26$  & \\
 $0.282$  &   $3.22\pm 0.26$  & \\
 $0.291$  &   $4.05\pm 0.66$  & \\
 $0.451$	&   $3.97\pm 0.19$  & \\
 $0.546$	&   $3.69\pm 0.37$  & \\
 $0.55$   &   $4.59\pm 0.36$  & \\
\hline\hline
\end{tabular}
\end{table}

It has been shown that  the decaying-$\Lambda$ term affects the photon
temperature evolution at the early epoch~\cite{rf:Kimura}. In this section, we study
about the consistency of D$\Lambda$CDM with recent temperature observations.

The temperature evolution is obtained from Eq.~\eqref{p14_g}.
Following the Stefan-Boltzmann's law, $\rho^{}_{\gamma} \propto T^4_{\gamma}$, 
 the relation between the photon temperature and the cosmological redshift is
 obtained as follows~\cite{rf:Puy,Nakamura2008}:
\begin{align}
 T^{}_{\gamma}(z) &= {T^{}_{\gamma0}}\left( 1 + z \right)\left(
					1+\frac{\Omega^{}_{\gamma\Lambda}}{\Omega^{}_{\gamma0}}\right)^{1/4}_{},
\label{eq:temp_evolv}
\\
\Omega^{}_{\gamma\Lambda} &\equiv
\begin{cases}
\alpha\left( \left( 1 + z\right)^{m-4}_{}-1 \right)  ~(m\ne 4),\\
-4{\Omega^{}_{\Lambda2}}{\ln{\left( 1+z \right)}}~ (m=4).
\end{cases}
\notag
\end{align}

From Eq.~(\ref{eq:temp_evolv}),  the
temperature-redshift relation at higher-$z$ approaches that of  S$\Lambda$CDM
model, $T^{}_{\gamma}\propto \left( 1+z \right)$. 
At lower-$z$,  the temperature evolution deviates from the proportional relation.

In the meanwhile, recently the temperatures at higher-$z$ are observed
 using the molecules such as 
the fine structure of 
excitation levels of the neutral or ionized
 carbon~\cite{Songaila1994,Ge1997,Srianand2000,Molaro2002,Cui2005} and
the rotational excitation of CO~\cite{Srianand2008,Noterdaeme2010,Noterdaeme2011} and
 Sunyaev-Zel'dovich (S-Z) effect~\cite{Battistelli2003, Luzzi2009}.
Furthermore, there are a lot of  temperature observations for the lower-$z$ owing to the S-Z effect as
shown in Table~\ref{tab:temp_obs}.
Therefore, we expect to constrain more severely the allowable  region on the
 $m-\Omega_{\Lambda2}$ plane from the new temperature  observations.

\begin{figure}[tb]
\begin{center}
 \includegraphics[width=.70\linewidth,keepaspectratio]{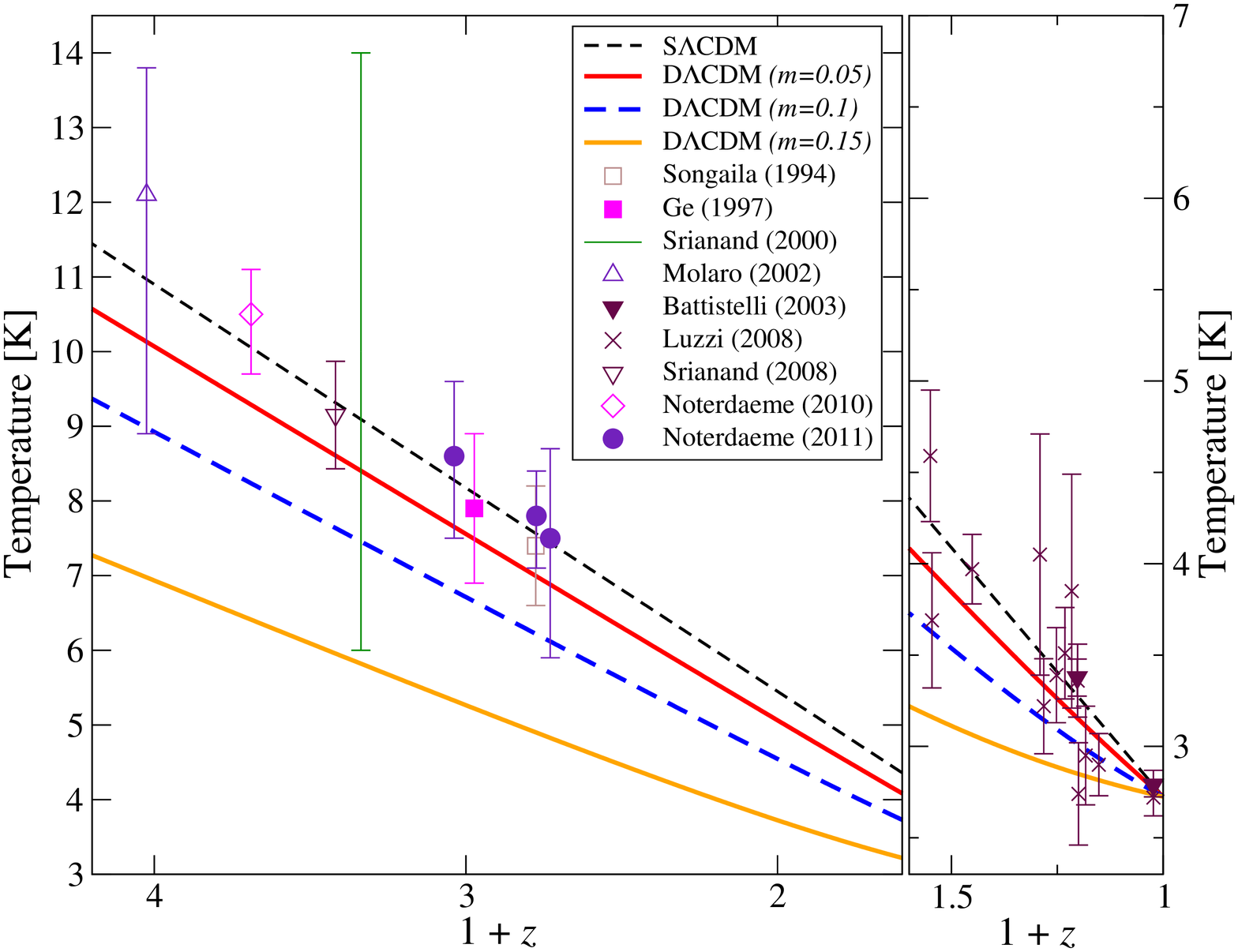}
\caption{Illustration of the temperature evolutions compared with CMB
 observations. The left panel is the results at $z>0.6$. The right  is
 same, but at $z<0.6$.}
\label{fig:dlcdm_temp}
\end{center}
\end{figure}

Figure~\ref{fig:dlcdm_temp} illustrates the temperature evolution in
S$\Lambda$CDM and  D$\Lambda$CDM model with
$\Omega_{\Lambda2}=10^{-2}$.  It is shown that the larger value of $m$
results in the lower-$T^{}_{\gamma}$.   When we adopt $m=0.15$,
the theoretical curve of $T_{\gamma}$ should be inconsistent with
observations even for  the temperature of  the largest uncertainty~\cite{Srianand2000}.
We can conclude that the deviation from S$\Lambda$CDM model might be small for $m<0.1$.

Using the  recent observations of the CMB temperature whose accuracy has been improved
quantitatively, we can put severe constraints on the temperature evolution in
D$\Lambda$CDM model.  We calculate $\chi^2_{}$-analysis as follows,
\begin{equation}
 \chi^{2}_{} = \sum_{i}^{N}\frac{ \left( T^{}_{\gamma}(z) -  T^{}_{obs, i}\right)^2_{}}
{\sigma^{2}_i},
\label{eq:chi2_temp}
\end{equation}
where $T(z)$ is theoretical temperature calculated
by Eq.(\ref{eq:temp_evolv}) and $T^{}_{obs, i}$ the observational data are shown in Table~\ref{tab:temp_obs}.
$\sigma_i$ is the uncertainty of the observation. $N=25$ is the number of the
observational data. 
Note that the result of Srianand et al.~\cite{Srianand2000} does not
have best-fit value.  Therefore, we assume the best-fit value is the
same as the mean value, $T_{\bf Srianand}=10.0\pm 4.0$~K.

In Fig.~\ref{fig:snIa_cntr},  we show the limits in the  $m-\Omega^{}_{\Lambda2}$
plane calculated by Eq.(\ref{eq:chi2_temp}). 
{
The shaded region indicates the parameter regions which should be
excluded and are from Eq.~\eqref{eq:condition_oml2}.
}
We can obtain the allowed parameter range as follows:
\begin{equation}
{
 \Omega_{\Lambda2} < 6.1\times10^{-4}~~\text{at 1$\sigma$ C.L.} ~\text{~and~}1.7\times10^{-3}~~\text{at 2$\sigma$ C.L.}}
\label{eq:limit_temp}
\end{equation}
Although temperature constraints give the upper bound, we can obtain the
parameter range which is more severe  than
that obtained in the previous temperature constraint~\cite{rf:Puy}.

Finally, we identify the influence of temperature observations at $z<0.6$
by S-Z effect. We also calculate the $\chi^2$-fitting
(Eq.~\eqref{eq:chi2_temp}) using observations at $z>0.6$ and $z<0.6$, separately.
As the results, we obtain the upper bound of $1\sigma~(2\sigma)$ C.L. as follows:
{
\begin{align} 
\Omega^{}_{\Lambda2} &< 2.8\times10^{-3}~(4.7\times 10^{-3})~~\text{by
 S-Z at } z < 0.6 \notag \\
\Omega^{}_{\Lambda2} &< 5.5\times10^{-4}~(1.9 \times 10^{-3})~~
 \text{by CI and CO at } z > 0.6 \notag.
\end{align}
}
This results are suggested that the temperature observations at a
higher-$z$ is important to constrain D$\Lambda$CDM model.


\section{Concluding Remarks}

We have investigated the consistency of the decaying-$\Lambda$ model
with both the $m-z$ relation of SNIa and CMB temperature.
We obtain the upper bound of our model parameters from both observations.
First we have constrained our model from SNIa: 
we can obtain the upper-bound of parameters,  which region is
wider compared with previous constraints~\cite{rf:Puy,Nakamura2008}.
From this result, a decaying-$\Lambda$ has minor effects on the cosmic
expansion at low-$z$ and its parameters cannot be well constrained .

On the other hand, we can obtain more severe parameter constraints from CMB temperature.
We acquire the limit of 
$\Omega^{}_{\Lambda2}$ as  {
 $\Omega_{\Lambda2} < 1.7\times10^{-3}$ at 2$\sigma$ C.L.
}
From this result, the temperature can become lower by a few percents. That
should affect the hydrogen recombination.
We note that we cannot get the bound of $m$. Because change in $\Omega^{}_{\Lambda2}$
cancels the effects of the variation of $m$.

Our parameter is mainly obtained by the observations at $z>2$.  
It has been suggested that the observations of the temperature at
higher-redshift is important to constrain the cosmological models with
 energy flow into photon~\cite{Nakamura2008}.
{
The upper-limit of $\Omega^{}_{\Lambda2}$ obtained from the present
constraints is still larger than the previous constraints of 
$\Omega^{}_{\Lambda2} < 1.7\times10^{-4}$ at $95.4\%$ C.L.~\cite{Nakamura2008}.
The analysis using the CMB temperature fluctuation of the recent
observation seems to be unable to give the tight constraints.}
We plan to perform more detailed analysis, for instance using CMB anisotropy
of the newest  WMAP data~\cite{WMAP7} .

\section{Acknowledgments}
This work has been supported in part by a Grant-in-Aid for Scientific
Research (18540279, 19104006, 21540272) of the Ministry of Education,
Culture, Sports, Science and Technology of Japan. 





\end{document}